\documentclass{article}
\usepackage{amsmath}
\usepackage{amssymb}

\input{tcilatex}

\begin{document}

\begin{center}
{\Large The gravitational collapse of a dust ball}

Trevor W. Marshall

CCAB, Cardiff University, 2 North Road, Cardiff CF10 3DY, UK
\end{center}

\textbf{Abstract } It is shown that the description of collapse given by the
classic model of Oppenheimer and Snyder fails to satisfy a crucial matching
condition at the surface of the ball. After correcting the model so that the
interior and exterior metrics match correctly, it is established that the
contraction process stops at the Schwarzschild radius, that there is an
accumulation of particles at the surface of the ball, and that in the limit
of infinite time lapse the density of particles at the surface becomes
infinite. A black hole cannot form. This result confirms the judgements of
both Einstein and Eddington about gravitational collapse when the collapse
velocity approaches that of light.

\section{Introduction}

The Relativistic Theory of Gravitation (RTG)\cite{logunov} may be considered
an example of a class of theories introduced by Rosen\cite{rosen}, known as 
\emph{bimetric theories}. Coexisting with the field, or Riemann metric%
\begin{equation}
ds^{2}=g_{\mu \nu }dx^{\mu }dx^{\nu }\quad ,
\end{equation}%
there is a space, or Minkowski metric%
\begin{equation}
d\sigma ^{2}=\gamma _{\mu \nu }dx^{\mu }dx^{\nu }\quad .
\end{equation}%
Unlike in General Relativity (GR), where the field is synonymous with the
geometry, we recognize that the field, as in the electromagnetic case, is
propagated through an underlying Minkowski space. In RTG there is a
preferred coordinate system, namely that for which the space metric is
Galilean. It may be called the \textit{inertial }system or frame, and its
corresponding field metric $g_{\mu \nu }$ gives rise to a gravitational
potential $\Phi ^{\mu \nu }=g^{\mu \nu }\sqrt{-g}$ whose divergence is zero.
The latter condition may be made covariant by requiring the covariant
divergence of the gravitational field in the Minkowski metric to be zero.
This is the coordinate system which Einstein\cite{gravwave} used in order to
derive his formula for the gravitational radiation emitted from a
time-varying quadrupole source.

In this article I shall show how the comoving coordinate frame introduced by
Tolman\cite{tolman}, and developed by Oppenheimer and Snyder\cite{oppsny}%
\cite{landau} (OS) to describe gravitational collapse, may be transformed to
the inertial, or harmonic frame. It is then possible to track the
trajectories of individual particles in the collapse of a dust ball, for
which the equation of state is simply $p=0$. We establish that the collapse
can go no further than the Schwarzschild radius, that the collapse to this
state takes an infinite time, and that the density of particles in the limit
becomes infinite at the surface of the ball. This confirms the description
given by Logunov\cite{logunov}, in which gravity changes from being
attractive to repulsive for certain high-density conditions. The description
of OS, namely "continued gravitational contraction", which may be considered
the ancestor of the contemporary "black hole", is now seen, in the light of
RTG, to be incorrect. It is more appropriate to call the process one of
gravitational \emph{\ compression; }the combination of an attraction of the
surface particles with a repulsion of the particles beneath the surface
produces an infinite density at the surface; presumably this balance of
purely gravitational forces will be modified by the action of nuclear forces
in an actual neutron star, so the state of infinite density is never
reached. But, in any case, contraction does not go beyond the Schwarzschild
radius, as I shall be able to show by tracking the trajectories of
individual particles. The results obtained here confirm the intuitions of
both Eddington\cite{edding} and Einstein\cite{einvbh}, as expressed in the
early period of the black-hole hypothesis. Particles which crossed the
"event horizon" would acquire velocities exceeding that of light, which runs
contrary to the prescription of Einstein's Special Theory of 1905; so black
holes are ruled out by that theory.

It may be the case that a modified Einstein-Hilbert equation incorporating a
nonzero cosmological constant, a natural and possibly even necessary feature
of RTG, will, as suggested in a recent article of Gerstein, Logunov and
Mestvirishvili\cite{gersetal}, stop the compression process after a certain
time and convert it into a cyclical, or pulsating, process. However, it is
not clear what will happen if, as seems plausible, nuclear forces intervene
before these extra gravitational terms can take effect.

Although RTG is a distinct theory, constructed in opposition to GR and
following ideas planted by earlier workers like Rosen\cite{rosen} and Fock%
\cite{fock}, it is possible to view the results obtained in the present
article as simply a way of transforming a long established solution of the
GR field equations, namely that of Tolman, into a more realistic coordinate
system. Previous attempts of this nature, starting from the Schwarzschild
solution, are those of Eddington and Finkelstein and of Kruskal, described
in the book of Landau and Lifshitz\cite{landau} (see also Hartle\cite{hartle}%
), but these made no attempt to penetrate into the interior of the
collapsing object.

\section{The Oppenheimer-Snyder metric}

The Oppenheimer-Snyder (OS) field metric for a dust ball may be written

\begin{equation}
ds^{2}=d\tau ^{2}-V^{2}dR^{2}-W^{2}\left( d\theta ^{2}+\sin ^{2}\theta d\phi
^{2}\right) \quad ,
\end{equation}%
where%
\begin{equation}
\frac{W}{2m}=\left( \sqrt{R^{3}}-\frac{3\tau }{4m}F\left( R\right) \right)
^{2/3},\quad \frac{V}{2m}=\frac{2m\sqrt{R}-\tau F^{\prime }\left( R\right) }{%
\sqrt{2mW}}\quad .
\end{equation}%
In this comoving metric the free-fall radial geodesics are simply $R=$%
constant, and the coordinate $\tau $ is the particle's proper time; in
particular the surface of the ball is specified by a fixed $R,$ which will
be put as $1$. The function $F\left( R\right) $, or rather the product $%
FF^{\prime }$, gives the mass distribution of the dust particles. For $R>1$
(the external region), we put%
\begin{equation}
F\left( R\right) =1\quad (R>1)\quad ,
\end{equation}%
giving zero density there, and for $R<1$ (the internal region), $F$ is left
arbitrary for the moment, but we note that, for the metric to be continuous
it has to satisfy%
\begin{equation}
F(1-)=1,\quad F^{\prime }\left( 1-\right) =0\quad .
\end{equation}%
At this point OS made a fatal error by choosing an $F$ which fails to
satisfy the second of these, so my choice of $F$ will constitute a corrected
version of OS. The cumulative mass distribution is given by the function%
\begin{equation}
M\left( R\right) =\int_{0}^{R}4\pi T^{00}\left( R^{\prime }\right) \sqrt{%
-g\left( R^{\prime }\right) }dR^{\prime }=mF^{2}\left( R\right) \quad .
\label{OSdensity}
\end{equation}

The harmonic coordinates\cite{fock}\cite{weinberg}\cite{logmesh} $\left(
t,r,\theta ,\phi \right) $ for this system are given by the solutions of%
\begin{equation}
\square t=0,\quad \square r=-\frac{2r}{W^{2}}\quad ,  \label{harm}
\end{equation}%
where the spherical d'Alembertian operator is given by%
\begin{equation}
\square =\partial _{\tau }^{2}-\frac{1}{V^{2}}\partial _{R}^{2}+\left( \frac{%
\dot{V}}{V}+\frac{2\dot{W}}{W}\right) \partial _{\tau }+\frac{1}{V^{2}}%
\left( \frac{V^{\prime }}{V}-\frac{2W^{\prime }}{W}\right) \partial
_{R}\quad ,  \label{dalembert}
\end{equation}%
and we use dot and prime to signify differentiation with respect to $\tau $
and $R$ respectively. The solution will be chosen to satisfy the asymptotic
conditions%
\begin{equation}
t\sim \tau ,\quad r\sim W\quad \left( \tau \rightarrow -\infty \right) \quad
,  \label{asym}
\end{equation}%
and in that case the evolution $R=$constant describes, in the asymptotic
region, the Newtonian collapse of a dust ball of uniform density. This may
be demonstrated from the relation between $r$ and $t$, which is%
\begin{equation}
r\left( R,t\right) \sim \left[ \frac{9m|t|^{2}F^{2}\left( R\right) }{2}%
\right] ^{1/3}\quad \left( t\rightarrow -\infty \right) \quad ,
\end{equation}%
which, combined with (\ref{OSdensity}), gives%
\begin{equation}
M\left( R\right) =m\left[ \frac{r\left( R,t\right) }{r\left( 1,t\right) }%
\right] ^{3}\quad ,
\end{equation}%
and shows that the mass contained in a ball of radius $r$ is proportional to
the ball's volume. The $t$-dependence of $r$ has, of course, been known
since Michell discovered it in the eighteenth century.

In the exterior region the solutions of (\ref{harm}) are\cite{logmesh}%
\begin{equation}
t=\tau -2\sqrt{2mW}+2m\ln \frac{\sqrt{W}+\sqrt{2m}}{\sqrt{W}-\sqrt{2m}}%
,\quad r=W-m\quad \left( R>0\right) \quad ,  \label{extsoln}
\end{equation}%
and the OS metric becomes, in this region,%
\begin{equation}
ds^{2}=\frac{r-m}{r+m}dt^{2}-\frac{r+m}{r-m}dr^{2}-\left( r+m\right)
^{2}\left( d\theta ^{2}+\sin ^{2}\theta d\phi ^{2}\right) \quad ,
\label{extmet}
\end{equation}%
which is the Schwarzschild metric, except that the Schwarzschild "radius" $r$
has been replaced by $r+m.$ We shall see that the solution of (\ref{harm})
may be continued into the interior region all the way to $R=0$, and that
this corresponds to $r=0$ for all $t.$ I deduce that it is this $r$, rather
than Schwarzschild's, which should be regarded as the true radius, a
conclusion which will be reinforced by making a closer examination of the
internal solutions of (\ref{harm}). It is clear that the above solution (\ref%
{extsoln}) gives, as $t$ goes to plus infinity, that $W$ approaches $2m$ and 
$r$ approaches $m$, and that these limits are reached at a finite value of
proper time 
\begin{equation}
\tau _{f}\left( R\right) =\frac{4m}{3}\left( \sqrt{R^{3}}-1\right) \quad
\left( R>1\right) \quad .
\end{equation}%
The fact that this is finite has led to the widespread, indeed almost
universally held, conclusion that a falling particle goes on to cross the
"event horizon" at $r=m$ for proper times greater than $\tau _{f},$ and
therefore to be swallowed by the black hole at $r=0$. I shall show that this
conclusion is incorrect.

All I have to do is demonstrate that the interior solutions of (\ref{harm})
give a limiting value $r_{f}\left( R\right) $ in $0<R<1$, with $r_{f}\left(
0\right) =0$ and $r_{f}\left( 1\right) =m$, together with a corresponding $%
\tau _{f}\left( R\right) $, for which $t$ goes to plus infinity. I have been
able to make the demonstration numerically for $r_{f}\left( R\right) $ with
a particular choice of the function $F\left( R\right) .$ However, for
general $F,$ the values of $\tau _{f}\left( R\right) $ may be obtained
simply by examining the characteristics of (\ref{harm}). These satisfy, for
both of these partial differential equations, the same pair of ordinary
differential equations%
\begin{equation}
\frac{d\tau }{dR}=\pm V\left( \tau ,R\right) =\pm \left( 2m\sqrt{R}-\tau
F^{\prime }\right) \sqrt{\frac{2m}{W}}\quad .  \label{chargen}
\end{equation}%
It is a simple matter to verify that the above value of $\tau _{f}\left(
R\right) $ satisfies this with the upper sign in $R>1$, where $F=1$ and $%
F^{\prime }=0$. In the interior region the upper-sign characteristic through
the point 
\begin{equation}
\tau _{f}\left( 1\right) =0  \label{charbound}
\end{equation}%
is the first one to be met by a geodesic $R=$constant coming from $\tau
=-\infty .$ The situation is illustrated in Figure 1, where I have plotted
this characteristic, in both exterior and interior regions, for the case $%
2m=1$, using in the interior the function $F$ of my previous article\cite%
{coldstar}, namely%
\begin{equation}
F\left( R\right) =R^{3/2}e^{3X/2},\quad X=1-R\quad (R<1)\quad .
\label{Fspec}
\end{equation}%
\FRAME{ftbpFU}{4.5792in}{3.7715in}{0pt}{\Qcb{The limit of physical
space-time with $2m=1$. $\protect\tau $ is the proper time of a dust
particle and $R$ is its comoving coordinate, so that $R=1$ indicates a
surface particle. A given dust particle, or in the exterior region a test
particle, moves along the abcissa $R=$constant, arriving at the boundary
curve after an infinite time $t.$}}{}{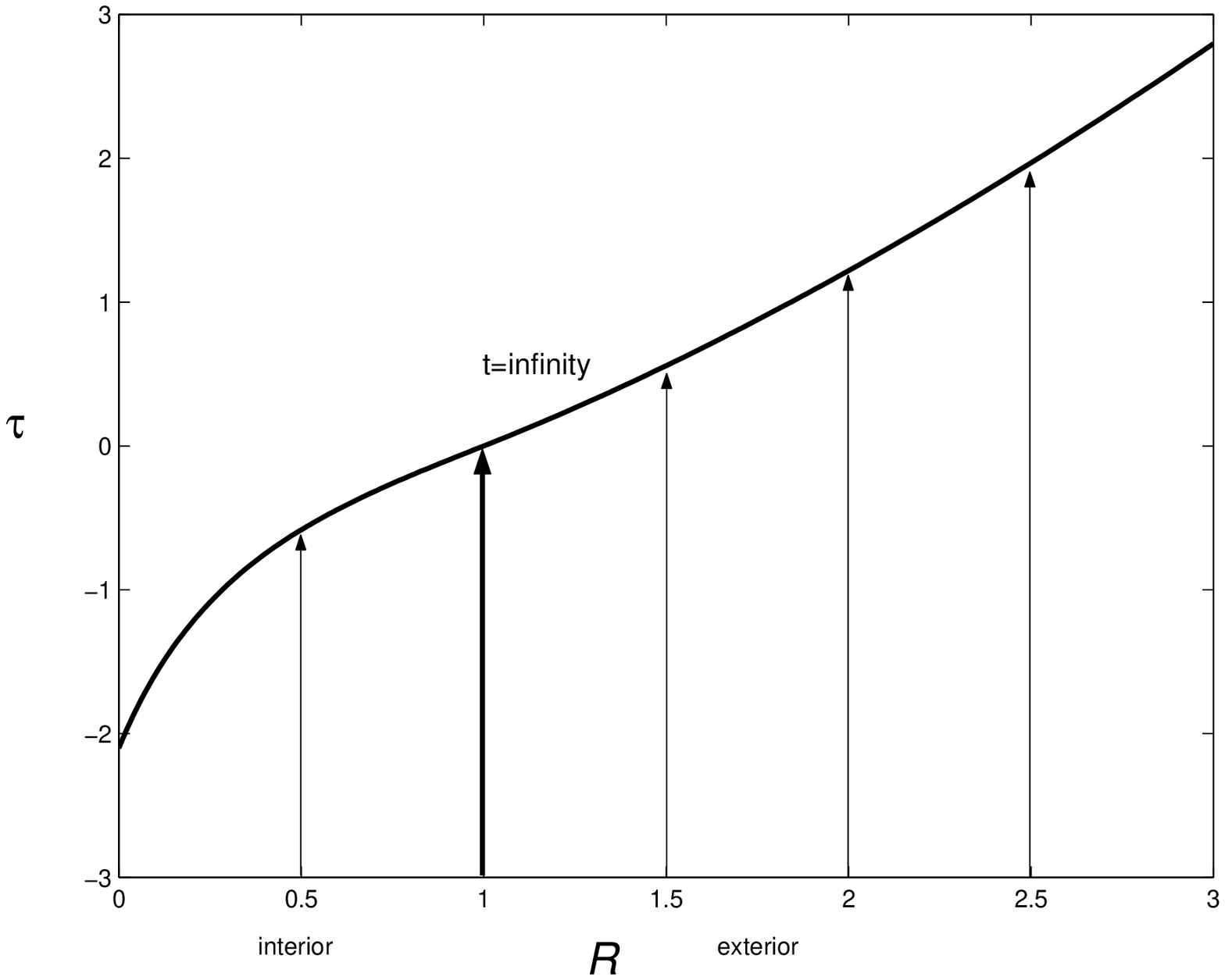}{\special{language
"Scientific Word";type "GRAPHIC";maintain-aspect-ratio TRUE;display
"USEDEF";valid_file "F";width 4.5792in;height 3.7715in;depth
0pt;original-width 6.762in;original-height 5.5642in;cropleft "0";croptop
"1";cropright "1";cropbottom "0";filename 'physreg.eps';file-properties
"XNPEU";}}

The figure represents the whole of space, that is $0<r<+\infty ,$ and the
whole of time, that is -$\infty <t<+\infty ;$ there are no singularities and
no trapped surfaces of the kind supposed by Penrose and Hawking\cite{penrose}%
. Such strange objects are to be found in the region beyond $t=+\infty $,
and belong in the realm of science fiction. The finite value of the proper
time $\tau _{f}$ is an indication that a falling particle, as it approaches
the boundary characteristic passing through the point $R=1,\tau =0$, suffers
an infinite gravitational red shift. We shall see a natural explanation for
this in the infinite surface density of the dust ball in the limit $%
t\rightarrow +\infty $.

The integration of (\ref{harm}) in the interior region may be best achieved
by changing to the characteristic coordinates\emph{\ }$\left( R,S\right) $,
defined by 
\begin{equation}
\tau =\tau \left( R,S\right) ,\quad \tau \left( 1,S\right) =-2mS,\quad
D_{R}\tau =V=\left( 2m\sqrt{R}-\tau F^{\prime }\right) \sqrt{\frac{2m}{W}}%
\quad .  \label{chareq}
\end{equation}%
The family of curves given by $S=$constant are the upper characteristics of (%
\ref{harm}), and in particular the boundary characteristic is $S=0$. In
terms of these coordinates the d'Alembertian (\ref{dalembert}) is
transformed by putting%
\begin{equation}
D_{R}=\partial _{R}+V\partial _{\tau }\quad ,
\end{equation}%
where $D_{R}$ is the derivative with respect to $R$ along the
characteristic, giving%
\begin{equation}
\square =-\frac{1}{V^{2}}D_{R}^{2}+\frac{2}{V}\partial _{\tau }D_{R}+\frac{1%
}{V^{2}}\left( \frac{V^{\prime }}{V}-\frac{2V}{W}\right) D_{R}+\frac{2}{W}%
\left( 1+\dot{W}\right) \partial _{\tau }\quad .
\end{equation}%
Note that $\partial _{\tau }$ does not commute with $D_{R},$ but that if we
write it as%
\begin{equation}
\partial _{\tau }=\Psi ^{-1}D_{S},\quad \Psi =\partial _{S}\tau \left(
R,S\right) \quad ,
\end{equation}%
then $D_{R}$ and $D_{S}$ commute. In the numerical procedure, described in
the next Section, there is no need to find the function $\Psi $ explicitly,
because we work with the differential operator $\partial _{\tau }$ rather
than $D_{S}.$ Although this integration is rather formidable, it may be
guaranteed that no singularity occurs in the whole physical region, because
the coefficients of the PDEs are nonzero and nonsingular, and the
determinant of the second-order coefficients is negative throughout, thereby
preserving their hyperbolic character.

Note that the characteristic curves we just introduced are also the null
geodesics of the (corrected) OS metric. They are the paths taken by outward
going light signals, and since the local light velocity is $V,$ their form
underlines the importance of the correction I made to the original OS
metric, insisting on $V$ being continuous at $R=1$. The time required for
such a signal to go from an internal point $(R_{1},S)$ to an external point $%
\left( R_{2},S\right) $ may be written as%
\begin{equation}
t\left( R_{2},S\right) -t\left( R_{1},S\right) =t\left( 1,S\right) -t\left(
R_{1},S\right) +G\left( R_{2},S\right) \quad ,
\end{equation}%
where $G$ is the travelling time in the exterior region, that is%
\begin{equation}
G\left( R_{2},S\right) =t\left( R_{2},S\right) -t\left( 1,S\right) \quad .
\end{equation}%
In view of the simple form (\ref{extmet}) of the exterior metric, this
latter integral may be simplified to give%
\begin{equation}
G\left( R_{2},S\right) =r\left( R_{2},S\right) -r\left( 1,S\right) +2m\ln 
\frac{r\left( R_{2},S\right) -m}{r\left( 1,S\right) -m}\quad .
\end{equation}%
Now, since $r(1,S)>m$ for all $t$, this result shows that any light signal
emitted from inside the ball eventually reaches the exterior region, and
this should lay to rest all preexisting ideas regarding trapped surfaces.
But note that I said "eventually"; owing to the infinite red shift suffered
at the boundary $R=1$, this travel time becomes infinite as $r\left(
1,S\right) $ approaches $m$, that is in the closing stages of the
compression process.

\section{Numerical integration}

\bigskip It is convenient to define the operator 
\begin{equation}
P=-V^{2}\square =D_{R}^{2}-2\xi W\partial _{\tau }D_{R}-2\xi ^{2}W\left( 1+%
\dot{W}\right) \partial _{\tau }+\left( \xi -\frac{\xi ^{\prime }}{\xi }%
\right) D_{R}\quad ,  \label{Pdal}
\end{equation}%
where $\xi =V/W$, that is, for the choice we made in (\ref{Fspec}) for $F$, 
\begin{equation}
\xi =R^{-1}\left( 1+\frac{3\tau }{4m}Xe^{3X/2}\right) \left( 1-\frac{3\tau }{%
4m}e^{3X/2}\right) ^{-1}\quad ,
\end{equation}%
and the harmonic coordinates then satisfy%
\begin{equation}
Pt=0,\quad \left( P-2\xi ^{2}\right) r=0\quad .  \label{harm1}
\end{equation}%
These PDEs must be integrated in $0<R<1,S>0$ with surface boundary
conditions at $R=1$%
\begin{equation}
t\left( 1,S\right) =-2mS-2\sqrt{2mW_{0}}+\ln \frac{\sqrt{W_{0}}+\sqrt{2m}}{%
\sqrt{W_{0}}-\sqrt{2m}},\quad r\left( 1,S\right) =W_{0}-m\quad ,
\end{equation}%
where%
\begin{equation}
W_{0}\left( S\right) =W\left( 1,S\right) =2m\left( 1+\frac{3S}{2}\right)
^{2/3}\quad ,
\end{equation}%
and also with the asymptotic condition (\ref{asym}) for $S\rightarrow \infty 
$ and with finite values at $R=0$.

Because $t\left( 1,S\right) $ becomes infinite at $S=0$, the integration
must be over a range $S\geq S_{1}>0$, and we may then convert each PDE into
a set of coupled ODEs in $R,$ starting from $t\left( 1,S\right) $ and $%
r\left( 1,S\right) $, for a discrete set of $N$ values of $S$ between $S_{1}$
and an upper limit $S_{2}$ for which the asymptotic values may be used.
Because $S$ is itself defined by the first-order ODE (\ref{chareq}), we
retain $\tau $ as the dependent variable, so there are a total of $3N$
coupled ODEs for $t,Dt$ and $\tau $. Fuller details, including the
asymptotic matching procedure, are given in the Appendix.

From the solutions $r\left( R,S\right) $ and $t\left( R,S\right) $ I
interpolated to obtain $r\left( R,t\right) $. Putting $r_{0}\left( t\right)
=r\left( 1,t\right) $, the relative position of a dust particle whose
position in the ball is indexed by $\left( R,\theta ,\phi \right) $, with $%
0<R<1$, may then be described by the coordinate%
\begin{equation}
u\left( R,r_{0}\right) =\frac{r\left[ R,t\left( r_{0}\right) \right] }{r_{0}}%
\quad \left( 0<R<1\right) ,\quad \quad .
\end{equation}%
I have plotted, in Figure 2, $u\left( R,r_{0}\right) $ against $R$ for
various values of $r_{0}$. In the early stage of collapse the slopes at both 
$R=0$ and $1$ increase as $r_{0}$ decreases, indicating that particles near
the centre move towards the surface and particles near the surface move
towards the centre. However, towards the end of the process the curve $%
u(R,r_{0})$ almost immediately crosses the curve $u\left( R,\infty \right) $
(the lower bold curve in Figure 2) near $R=1,$ and in the limit $%
r_{0}\rightarrow m$ approaches the upper bold curve. The latter has both
zero slope and zero curvature at $R=1$, indicating that the particle
distribution in the final state has an infinite density at the surface.

\FRAME{ftbpFU}{4.8308in}{3.7005in}{0pt}{\Qcb{Evolution of the particle
distribution $u\left( R\right) $. $u$ is the position of a given particle
relative to the surface, indexed by its comoving coordinate $R$. The lower
bold curve gives $u(R)$ in the limit $r_{0}\rightarrow \infty $, that is $%
t\rightarrow -\infty $, while the upper bold curve gives $u\left( R\right) $
for $r_{0}=m$, that is $t\rightarrow +\infty $. The lighter curves give $%
u\left( R\right) $ for the values \emph{(i) }$r_{0}=2m$ \emph{(ii) }$%
r_{0}=1.4m$ \emph{(iii) }$r_{0}=1.1m$.}}{}{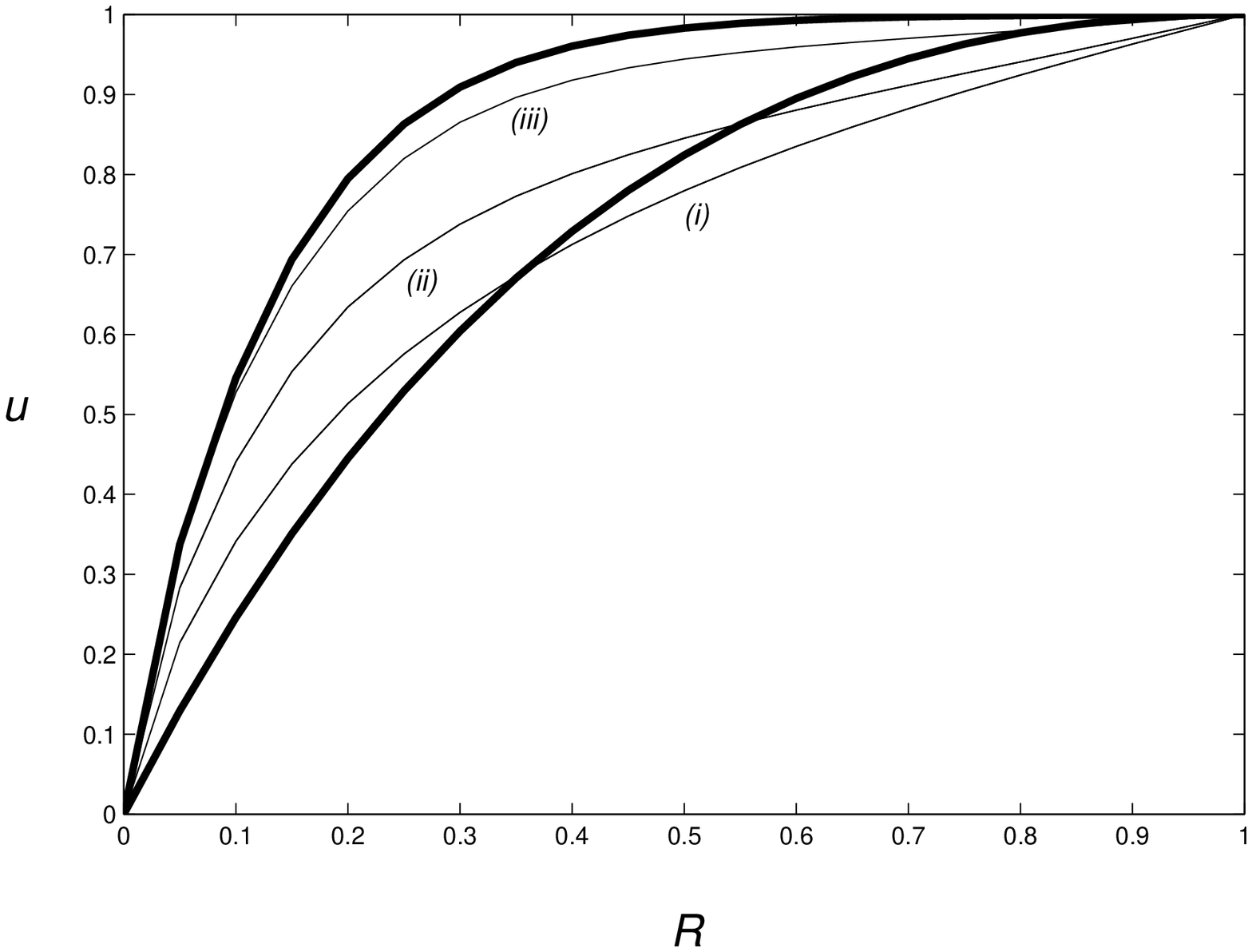}{\special{language
"Scientific Word";type "GRAPHIC";maintain-aspect-ratio TRUE;display
"USEDEF";valid_file "F";width 4.8308in;height 3.7005in;depth
0pt;original-width 7.1131in;original-height 5.4379in;cropleft "0";croptop
"1";cropright "1";cropbottom "0";filename 'collevol1.eps';file-properties
"XNPEU";}}

\section{Conclusion}

A surprising feature of Figure 2 is that the density of dust particles,
which started off uniform in the Newtonian region of $t$, becomes infinite
at the surface of the ball as $t\rightarrow +\infty $; this may be
considered as confirming Eddington's intuition\cite{edding} that something
intervenes to prevent the "absurdity" (to quote Eddington) of the ball
collapsing to a point. We see that the finite radius of the end state is a
consequence of a purely gravitational field, without the need for any other
forces to counter the gravitational attraction. Indeed, if we commit the
heresy of speaking of gravitational forces, we may say that they turn from
being attractive to repulsive in the high-density region; it is the
accumulation of gravitational energy inside the ball which prevents it from
collapsing to a point.

Such heresy is at least hinted at by the favoured status afforded here to
the harmonic coordinate frame, often referred to as a "gauge". But is it
possible to put the results I have established into a coordinate-free form?
My central result is the form of the boundary characteristic given by (\ref%
{chargen}) and (\ref{charbound}). This makes no reference to the harmonic
frame, even though I went via that frame to establish it, so a
coordinate-free derivation may be possible.

I would, nevertheless, argue that use of the harmonic frame advances our
understanding of the collapse, and especially of the stable final state,
which was declared nonexistent in the OS analysis. There was a mystery about
the infinite red shift suffered by a free-falling exterior particle as it
approaches $r=m$. Now, in the light of the discovery that particles inside
the ball behave in a similar manner, such behaviour becomes less mysterious.
We merely have to find an explanation for the build up of density at the
surface, and that, expressed in suitable coordinates, is to be found in the
Einstein-Hilbert field equations themselves. It may be that some change of
frame could transform away such an infinite density so that it becomes
finite, but the theory of gravity which motivated the analysis of this
article\cite{logunov}, and also the earlier statement of opposition to
Strong Equivalence by Fock\cite{fock}, seem to offer substantial support to
the assertion that the harmonic, or inertial, coordinate frame should have a
privileged status.

Finally I remark that the harmonic frame was first accorded such a status in
Einstein's pioneer article on gravitational waves\cite{gravwave}. The
subsequent history\cite{kennefick} of gravitational waves is a process of
repeated assertion and denial of the Strong Equivalence Principle that all
frames are equivalent, beginning perhaps with Eddington's observation\cite%
{edding2} that, \emph{if we allow arbitrary coordinate transformations, the
waves may be made to appear and disappear at will; gravitational waves
travel at the speed of thought.} The converse of this statement, of course,
is that if we do not allow such transformations, but instead insist on a
privileged frame, such waves do exist and are worth searching for.

\section{Appendix}

The values of $t$ and $r$ in the asymptotic region may be obtained in terms
of the variables $\left( R,v\right) $, where%
\begin{equation}
v^{3}=e^{-3X/2}-\frac{3\tau }{4m},\quad X=1-R\quad .
\end{equation}%
Then the operator $P$, defined by (\ref{Pdal})$,$ becomes%
\begin{gather}
P=\partial _{R}^{2}+\frac{\eta }{v^{2}}\partial _{R}\partial _{v}+\left( 
\frac{\eta ^{2}}{4v^{6}}-\frac{\xi ^{2}R^{2}e^{2X}}{4v^{2}}\right) \left(
v\partial _{v}\right) ^{2}-\left( \frac{3\xi RXe^{2X}}{4v}+\frac{\eta X}{%
4Rv^{2}}-\frac{\eta ^{2}}{2v^{5}}\right) \partial _{v}  \notag \\
+\left( \frac{X^{2}+1}{RX}+\frac{5\eta }{2v^{3}}\right) \partial _{R}-\frac{%
\eta X}{Rv^{3}\xi }\partial _{R}\left( \frac{\eta }{X}\right) \left(
\partial _{R}-\frac{vX}{2R}\partial _{v}\right) \quad ,  \label{dal2}
\end{gather}%
where%
\begin{equation}
\xi =\frac{X}{R}+\frac{\eta }{v^{3}},\quad \eta =e^{-3X/2}\quad .
\end{equation}

By expanding this in inverse powers of $v$ and substituting in (\ref{harm1}%
), one obtains the first few terms in the asymptotic series of $t$, with the
first term given by the Newtonian limit, that is%
\begin{equation}
\frac{t}{2m}\sim -\frac{2}{3}v^{3}+\sum_{n=0}^{\infty }t_{n}\left( R\right)
v^{1-n}\quad ,
\end{equation}%
with%
\begin{eqnarray}
t_{0} &=&-\frac{1}{2}R^{2}e^{2X}-\frac{3}{2},\quad t_{1}=\frac{2}{3}%
e^{-3X/2}\quad ,  \notag \\
t_{2} &=&\frac{91}{40}-\frac{1}{4}R^{2}e^{2X}-\frac{1}{40}R^{4}e^{4X}\quad ,
\notag \\
t_{3} &=&\frac{1}{2}e^{-3X/2}+\frac{3}{2}\left( 13-6X+X^{2}\right)
e^{X/2}-20\quad ,  \notag \\
t_{4} &=&\frac{571}{672}-\frac{91}{480}R^{2}e^{2X}+\frac{1}{160}R^{4}e^{4X}+%
\frac{1}{3360}R^{6}e^{6X}\quad ,  \notag \\
t_{5} &=&\frac{5}{6}t_{3}+\frac{41}{120}\left( e^{-3X/2}-1\right)
+t_{51}-t_{52}+\frac{e^{-X}}{R}t_{53}+t_{52}\left( 0\right) -t_{53}\left(
0\right) \quad ,  \notag \\
t_{6} &=&\frac{1}{15}+\frac{3}{2}\left( 9-4X+X^{2}\right) e^{-X}-\frac{40}{3}%
e^{-3X/2}+\frac{1}{6}e^{-3X}\quad .
\end{eqnarray}%
where%
\begin{equation*}
t_{51}=\frac{1}{6}\int_{0}^{X}\frac{XU^{2}}{R}t_{3}\quad ,
\end{equation*}%
\begin{equation}
t_{52}=\frac{5}{16}\int_{X}^{1}\eta U^{4}\quad ,
\end{equation}%
and%
\begin{equation}
t_{53}=\frac{1}{5}\int_{X}^{1}\eta U^{5}\quad .
\end{equation}%
Note that the latter three integrals may all be given in terms of elementary
functions, though they are rather complicated.

The corresponding series for $r$ is%
\begin{equation}
\frac{r}{2m}=e^{X}R\left[ v^{2}+\sum_{n=0}^{\infty }r_{n}\left( R\right)
v^{-n}\right] \quad .
\end{equation}%
with the coefficients%
\begin{eqnarray}
r_{0} &=&\frac{1}{4}R^{2}e^{2X}-\frac{3}{4},\quad
r_{1}=r_{2}=r_{4}=r_{5}=r_{7}=0\quad ,  \notag \\
r_{3} &=&20-\frac{1}{2}e^{-3X/2}-\frac{3}{2}\left( 13-6X+X^{2}\right)
e^{X/2}\quad ,  \notag \\
r_{6} &=&\frac{5}{6}-\frac{15}{4}\left( 9-4X+X^{2}\right) e^{-X}+\frac{100}{3%
}e^{-3X/2}-\frac{5}{12}e^{-3X}\quad ,  \notag \\
r_{8} &=&-\frac{2403}{8}-\frac{27}{8}e^{X}\left(
X^{4}-9X^{3}+40X^{2}-89X+89\right)  \notag \\
&&+\frac{3}{8}e^{-X}\left( X^{2}+X+1\right) +\frac{3}{8}e^{2X}\left(
X-1\right) ^{2}+60e^{X/2}\left( X^{2}-5X+10\right)  \notag \\
r_{9} &=&\frac{16}{675}+\frac{20}{9}e^{-3X/2}-\left( 6X^{2}-\frac{96}{5}X+%
\frac{1158}{25}\right) e^{-5X/2}  \notag \\
&&+\frac{400}{9}e^{-3X}-\frac{10}{27}e^{-9X/2}
\end{eqnarray}%
In each coefficient of both expansions two constants of integration had to
be specified. They were determined from the continuity condition at $R=1$
and the finiteness condition at $R=0.$ The derivatives of $t_{n}$ at $R=1$
are fixed by this procedure, and we find that\footnote{%
Note that this result is at variance with the assumption of continuity for $%
t_{R}$ made by me in Ref\cite{coldstar}. Essentially, we must replace this
condition by one of finiteness at $R=0$. But I stress that the condition
imposed by Ref\cite{oppsny} on the interior metric in the entire region $R<0$
cannot be satisfied by any solution with an acceptable degree of continuity
at the surface.}%
\begin{equation}
t_{0}^{\prime }=t_{2}^{\prime }=t_{3}^{\prime }=t_{4}^{\prime
}=t_{6}^{\prime }=0,\quad t_{1}^{\prime }=1,\quad t_{5}^{\prime }=\frac{2}{5}%
\quad \left( R=1\right) \quad .
\end{equation}%
Rather remarkably, although the asymptotic series breaks down for $X=O\left(
v^{-3}\right) $, a match with the solution in this boundary layer confirms
the above values, and gives the additional ones%
\begin{eqnarray}
t_{6}^{\prime } &=&0,\quad t_{7}^{\prime }=\frac{68}{105},\quad
t_{8}^{\prime }=-0.0616\ldots ,\quad t_{9}^{\prime }=1\quad ,  \notag \\
t_{10}^{\prime } &=&0.6314\ldots ,\quad t_{11}^{\prime }=-2.3253\ldots \quad
.
\end{eqnarray}%
The corresponding procedure for the $r$ coefficients gives $r_{n}^{\prime
}=0 $ for $n\leq 9$, and the boundary layer matching not only confirms
these, but shows that $r_{10}^{\prime },r_{11}^{\prime },r_{12}^{\prime }$
are also zero. I have been emboldened to conjecture that the partial
derivative of $r$ with respect to $R$ is zero for all $v>1$.

The numerical integration procedure I have described requires a knowledge of
not only $t$ and $r$ at $R=1,$ but also their $R$-derivatives there$.$ For $r
$ the conjecture I just described may be tested by moving the lower limit of
the region of integration close to $S=0$, and when this was done there was
no noticeable tendency for the solution to diverge as the limit $R=0$ was
approached; of course rounding errors always allow the divergent solution to
enter at some stage, but this tendency seems to be constant over the whole
range of $S.$ In the case of $t$, all we can do is include all the
asymptotic coefficients we have, in order to approximate the initial value
of its $R$-derivative. Such a procedure gives a stable solution down to a
dustball radius $r_{0}$ of around $1.1m,$ and thanks to our rather precise $r
$-solution we already know what happens for infinite $t$, that is $r_{0}=m$.

\end{document}